\documentclass[aps,prl,twocolumn,groupedaddress,showpacs,floatfix]{revtex4-1}

\usepackage{dcolumn}
\usepackage{amsmath}
\usepackage{amssymb}
\usepackage{graphicx}
\usepackage{bm}
\usepackage[T1]{fontenc}
\usepackage{color}
\usepackage{url}
\usepackage[bf]{subfigure}
\usepackage{rotating}

\usepackage{soul}
\usepackage{scalefnt}

\newcommand\strength{S_i = \sum_{j \in r_i} {w_{ij}}}
\newcommand\stress{SC_u = \sum_{ij}{\sigma (i,u,j)}}
\newcommand\betweenness{B_u = \sum_{ij} \frac{\sigma (i,u,j)}{\sigma (i,j)}}
\newcommand\modularity{Q = \frac{1}{2m} \sum_{i,j} \left( A_{ij} - \frac{k_i k_j}{2m}\right) \delta(c_i,c_j) }

\begin{document}

\title{Minimal paths between communities induced by geographical networks}

\author{Henrique Ferraz de Arruda}
\email{h.f.arruda@gmail.com}
\affiliation{Institute of Mathematics and Computer Science at S\~ao Carlos, University of S\~ao Paulo, S\~ao Carlos, SP, Brazil}
\author{C\'esar Henrique Comin}
\email{chcomin@gmail.com}
\author{Luciano da Fontoura Costa}
\email{ldfcosta@gmail.com}
\affiliation{Institute of Physics at S\~ao Carlos, University of S\~ao Paulo, S\~ao Carlos, SP, Brazil}

\pacs{89.75.Fb}{Structures and organization in complex systems}

\begin{abstract}
In this work we investigate the betweenness centrality in geographical networks and its relationship with network communities. We show that
vertices with large betweenness define what we call characteristic betweenness paths in both modeled and real-world geographical networks. We define a geographical
network model that possess a simple topology while still being able to present such betweenness paths. Using this model, we show that such paths represent
pathways between entry and exit points of highly connected regions, or communities, of geographical networks. By defining a new network, containing information
about community adjacencies in the original network, we describe a means to characterize the mesoscale connectivity provided by such characteristic betweenness
paths.
\end{abstract}

\maketitle

\setcounter{secnumdepth}{1}
\section{Introduction}
\label{introduction}

As a consequence of their potential to represent real systems, complex networks have become the subject of growing interest in recent years
\cite{Strogatz2001Exploring}. Some examples of systems that have been studied using complex networks are electric power grids \cite{Motter2013power}, airline 
routes \cite{Lordan2014airline}, the World-Wide Web \cite{Albert2002web}, among others \cite{Newman2003Networks}. Networks whose vertices  are embedded in a
space with well-defined coordinates and their connectivity is related to the distance between the vertices are named geographical networks
\cite{barthelemy2011spatial}. Examples of geographical networks include street, airline, and  neural networks. Models to generate artificial networks with
geographic dependency have been created \cite{Boccaletti2006Complex} in order to better understand the real systems.

The spatial influence on the connectivity gives rise to many interesting characteristics \cite{Costa2007Characterization}. In a previous work about the
influence of gene expression on neuronal characteristics \cite{arruda2015framework}, we computed the betweenness centrality \cite{Freeman1977Betweenness} of
neuronal networks. While visualizing those neuronal networks we observed that the vertices having the highest betweenness centrality values were connected together
creating chains of vertices, which here we call betweenness paths. The presence of betweenness paths in the neuronal network suggests a two-level
hierarchy of the system structure, where the betweenness paths form the main pathways for information flow and the remaining paths distribute this information
locally. Such hierarchical organization has also been observed for road networks  \cite{bigotte2010integrated, crucitti2006urban, crucitti2006urban2,
scellato2006backbone, yerra2005emergence}. In \cite{eppell2001four} its argued that road networks should be planned in different scales, according to their
purpose, function or management policies. Such scales range from the entire city level down to local streets used for property access or pedestrian movement.
According to \cite{lammer2006scaling, chowell2003scaling} the traffic in road networks, modeled by applying the betweenness centrality, and the movement of
people in a city have also been shown to be power-law distributed, meaning that a few paths concentrate most of the city traffic.

Given that a hierarchical structure of pathways has been observed in a range of systems, it is interesting to study the reasons why they are formed, their
importance for the correct operation of the system, their influence in the network flow, and in which kind of geographical networks the betweenness paths can be
found. To better understand the hierarchical structures, it is necessary to study how the network topology is inducing those paths. Furthermore, the
knowledge about betweenness paths and their characteristics can be used for better planning the system.

In this work, we study street networks composed by two hierarchical levels: a main structure forming the backbone of the network, and secondary connections
supported by such backbone. We considered networks generated from geographic models and real-world street networks. From these networks, we identified connected
vertices with high betweenness centrality. We observed that these vertices provide a good covering  of the topology, in the sense that any network vertex is
aways near at least one or more characteristic betweenness paths. We found that such paths tend to be related with the community structure
\cite{Newman2006Modularity} of the network. So, to better understand the characteristic betweenness paths and its relationship with communities, we created a
random geographic model and used community detection to divide the space in well-defined regions. In order to study the relationship between such communities
and their geographical positions, we created a sub-summed community network, where each vertex is a different community and the connection weights are based on
the amount of shortest paths crossing the community border.

This paper is organized as follows. Section~\ref{concepts} presents the basic network theory concepts that are used in this paper. In Section~\ref{paths} 
we discuss betweenness paths in network models and real cases. In Section~\ref{model} we create a network model to better understand the betweenness paths. In
Section~\ref{community}, we describe the relationship between geographic and topological measurements.

\section{Basic concepts}
\label{concepts}

There are many ways to compute network centralities, each being developed to evaluate different network characteristics \cite{Boccaletti2006Complex}.

In order to locally quantify the network centrality, it is possible to measure the degree $k_i$ for each vertex $i$. Furthermore, in weighted networks, we can
consider the edges to define a measurement called strength \cite{Costa2007Characterization}, which is computed for each vertex $i$ as 
\begin{equation}
\label{eq.1}
\strength,
\end{equation}
where $r_i$ is the set of neighbors of vertex $i$ and $w_{ij}$ is the weight of the edge connecting $i$ to $j$. 

The network centrality can also be computed globally, that is, considering the whole network structure. One important global centrality measure is the stress
centrality ($SC_u$) \cite{Shimbel1953Stress}. This measure is defined for each vertex $u$ as
\begin{equation}
\label{eq.2}
\stress,
\end{equation}
where $\sigma(i, u, j)$ represents the number of shortest paths between the vertices $i$ and $j$ crossing the vertex $u$. If one vertex is removed from the
network it can affect the stress of all other vertices.

Another important global measurement is the betweenness centrality, which also considers the amount of shortest paths crossing a vertices
\cite{Freeman1977Betweenness}. This measure is defined as
\begin{equation}
\label{eq.3}
\betweenness,
\end{equation}
where $\sigma(i, u, j)$ represents the number of shortest paths between the vertices $i$ and $j$ crossing the vertex $u$ and $\sigma(i, j)$ is the total amount
of shortest paths between $i$ and $j$. The betweenness centrality can be implemented using a fast algorithm \cite{Brandes2001Betweeness, Brandes2008Social},
where the betweenness centrality is computed and the result is used to compute the stress centrality.

Betweenness centrality is frequently used to study flows in networks \cite{newman2005measure, borgatti2005centrality, freeman1991centrality} and can be used in
several applications. In social networks it can be described in terms of its message communications, considering that each message take the shortest path
\cite{Newman2010Networks}. In power grid networks, this measure can quantify the load of each vertex \cite{Motter2002Cascade}. It is also used in the
cascading-failure dynamics \cite{mirzasoleiman2011cascaded}.

In addition to the centrality measures, another important concept studied in complex networks theory is the community structure of the system. Where a community
is defined as a set of vertices densely interconnected, and with few connections to vertices outside this set.  This property can be found in many real networks such
as  social networks \cite{arenas2004community}, metabolic networks \cite{guimera2005functional}, and in the structure of cities \cite{zhong2014detecting}. In city networks, those communities are shown to be subdivisions of urban structure emerging urban elements and their interactions \cite{zhong2014detecting}.

A common approach to quantify how well-defined are the communities in the network is the modularity measure \cite{Newman2006Modularity}. The modularity is represented as
\begin{equation}
\label{eq.4}
\modularity,
\end{equation} 
where $m$ is the number of network edges, $A$ is the network adjacency matrix, $k_i$ and $k_j$ are the vertex degree $i$ and $j$, respectively, $c_i$ and $c_j$
are the communities of vertex $i$ and $j$, respectively, and $\delta(c_i,c_j)$ is $0$ if $c_i \neq c_j$ or $1$ if $c_i = c_j$.

Network communities can be detected through the optimization of modularity. For a comprehensive review of modularity optimization methods, please refere to 
\cite{fortunato2010community}. In this work we use the fast greedy \cite{Clauset2004FastGreedy} algorithm to detect the communities.

\section{Characteristic betweenness paths}
\label{paths}

In this section we define and analyze the characteristic betweenness paths. We compute the betweenness centrality of all the vertices. Most of these
vertices are connected, creating chains of vertices. Thus, most of them have degree two. We define a betweenness path as a sub-network considering only the the highest betweenness centrality values. In other words we selected only vertices with betweenness centrality higher or equal than a threshold and created a
sub-network composed by such vertices. In order to do so, we compute the betweenness centrality of the vertices in a network and divide them into two groups. Vertices having betweenness larger than a threshold $T$ are called \emph{characteristic vertices}, while the rest of vertices in the network are called \emph{secondary vertices}.

We will show four networks in which the characteristic paths can be observed. Two of these networks are artificially generated, and the other two are street
networks. In the first model, known as Waxman \cite{Waxman1988Connections}, the vertices are organized randomly in a plane. For each vertex, the probability to connect with all other vertices follows an exponential decay of the distance between them. In the second model, known as random geometric graph
\cite{Penrose2003Random}, the vertices are randomly organized and the connections are created only when the geographical distance is lower than a fixed value. We compared the network models with two real street networks: from San Joaquin County in USA and Oldenburg city in Germany \cite{Brinkhoff2002Cities}.

\begin{figure}
\center
\includegraphics[scale=0.45]{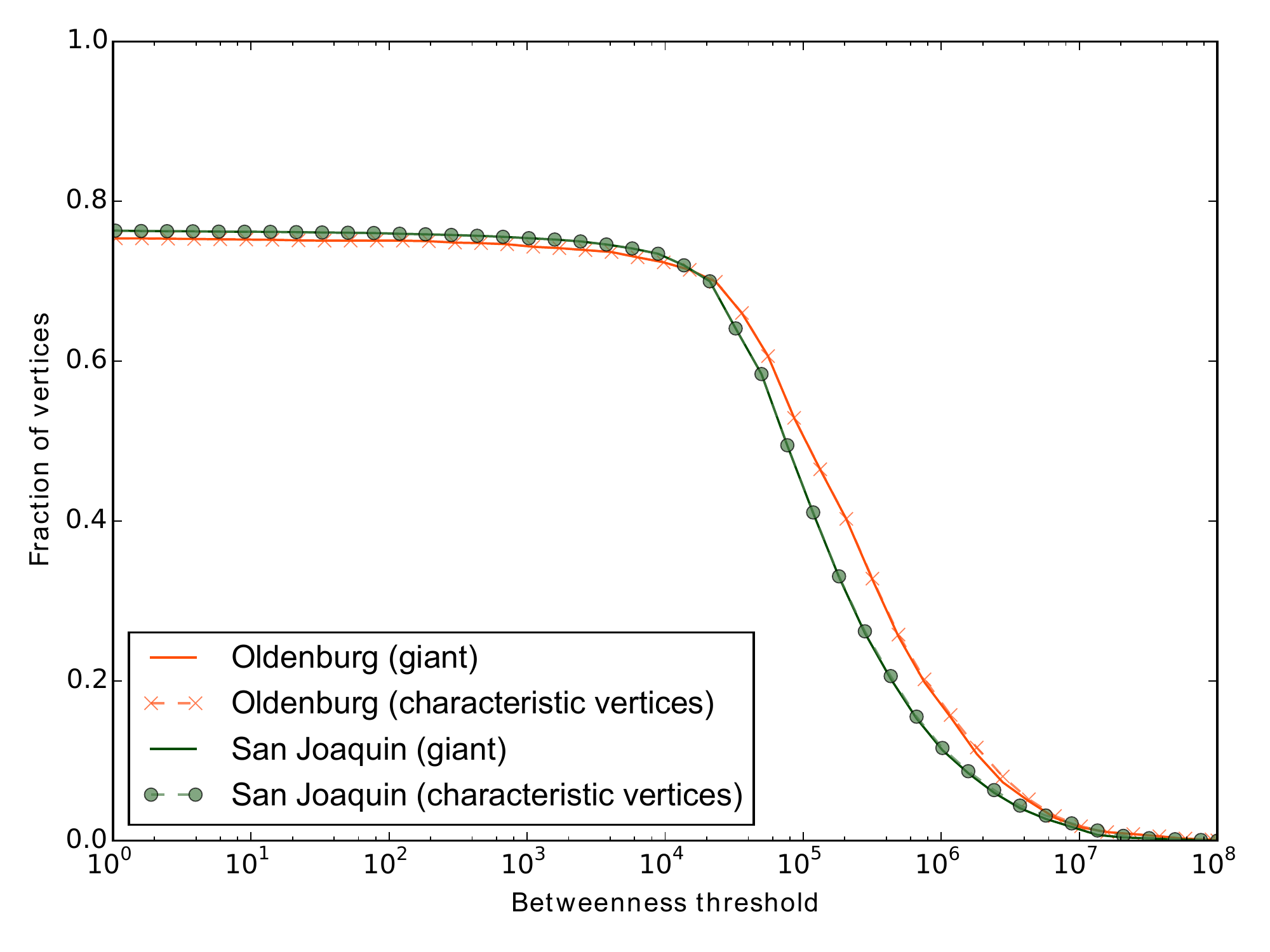}
\caption{(Color online) The plot shows the fraction of characteristic vertices against the betweenness threshold. For each network we also show the number of characteristic vertices
in the giant component.}
\label{fig.thresh}
\end{figure}

In order to find a reasonable value for the threshold, we computed the fraction of characteristic vertices according to some selected values of betweenness threshold. The fraction is found by dividing the number of characteristic vertices by the size of the network. We perform these tests using the street networks, the results can be seen in Figure~\ref{fig.thresh}.  Moreover, we also plot in Figure~\ref{fig.thresh} the largest connected component (giant component) of characteristic vertices. The majority of the characteristic vertices are included in the giant component for all different threshold values. Furthermore, the curves are similar for both cities. There is no indication of a proper threshold for defining the characteristic vertices. Therefore, we empirically selected a threshold  for which $20\%$ of the network vertices become characteristic vertices. If we consider the subnetwork formed by such vertices and their respective connections it is possible to show that for each analyzed network an interesting structure is obtained. Such structure contains a chain-like connectivity between the characteristic vertices, meaning that the majority of vertices have degree two. Such chain-like structure contains most of the flow in the network, represented by the betweenness, and can be seen as the backbone of the original network. The secondary vertices rely upon this backbone for proper communication with the rest of the system. The characteristic betweenness paths were also found in other works, but their origins were not analyzed \cite{crucitti2006urban, crucitti2006urban2, lammer2006scaling}.

\begin{figure*}
\center
\subfigure[]{\includegraphics[scale=0.41]{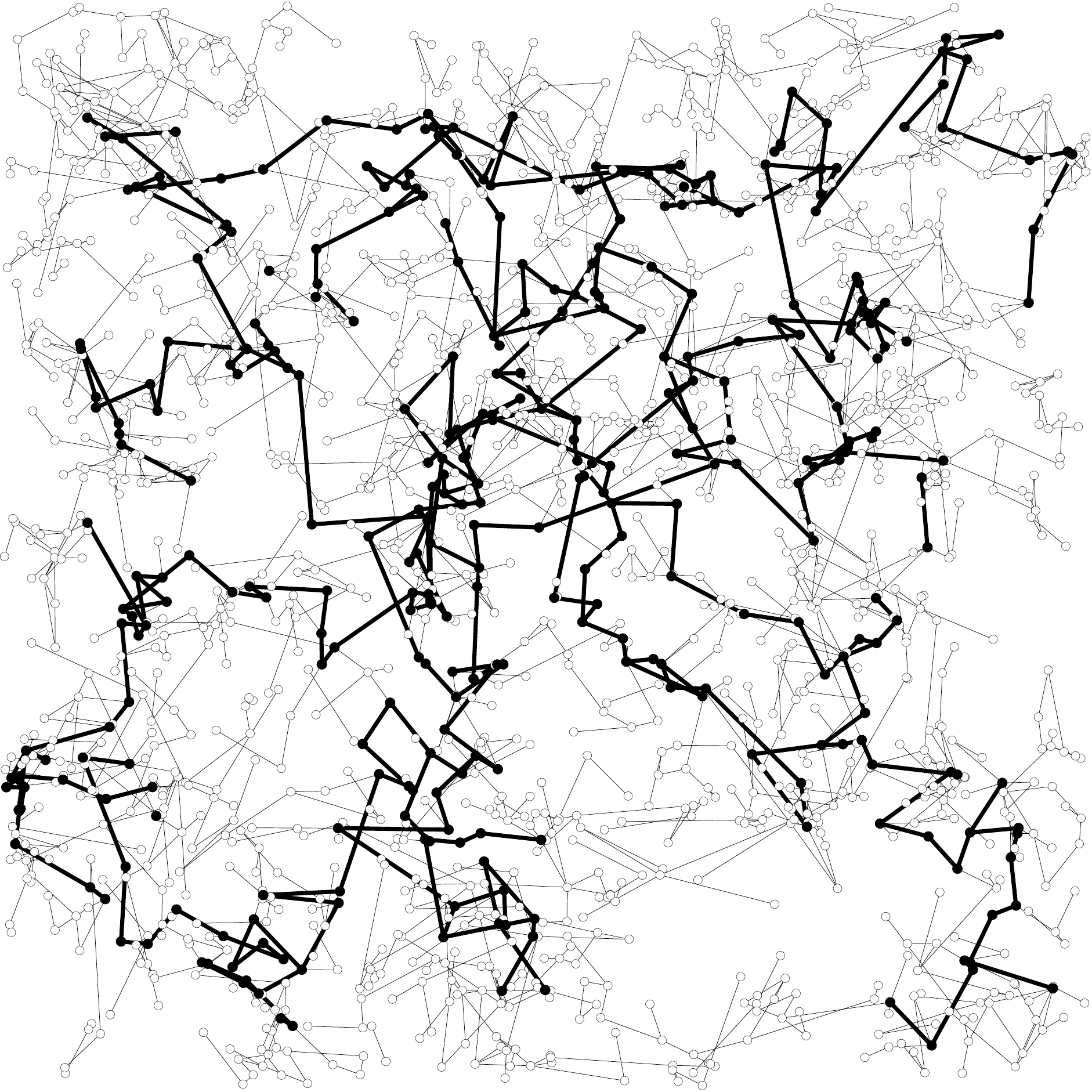}}
\subfigure[]{\includegraphics[scale=0.41]{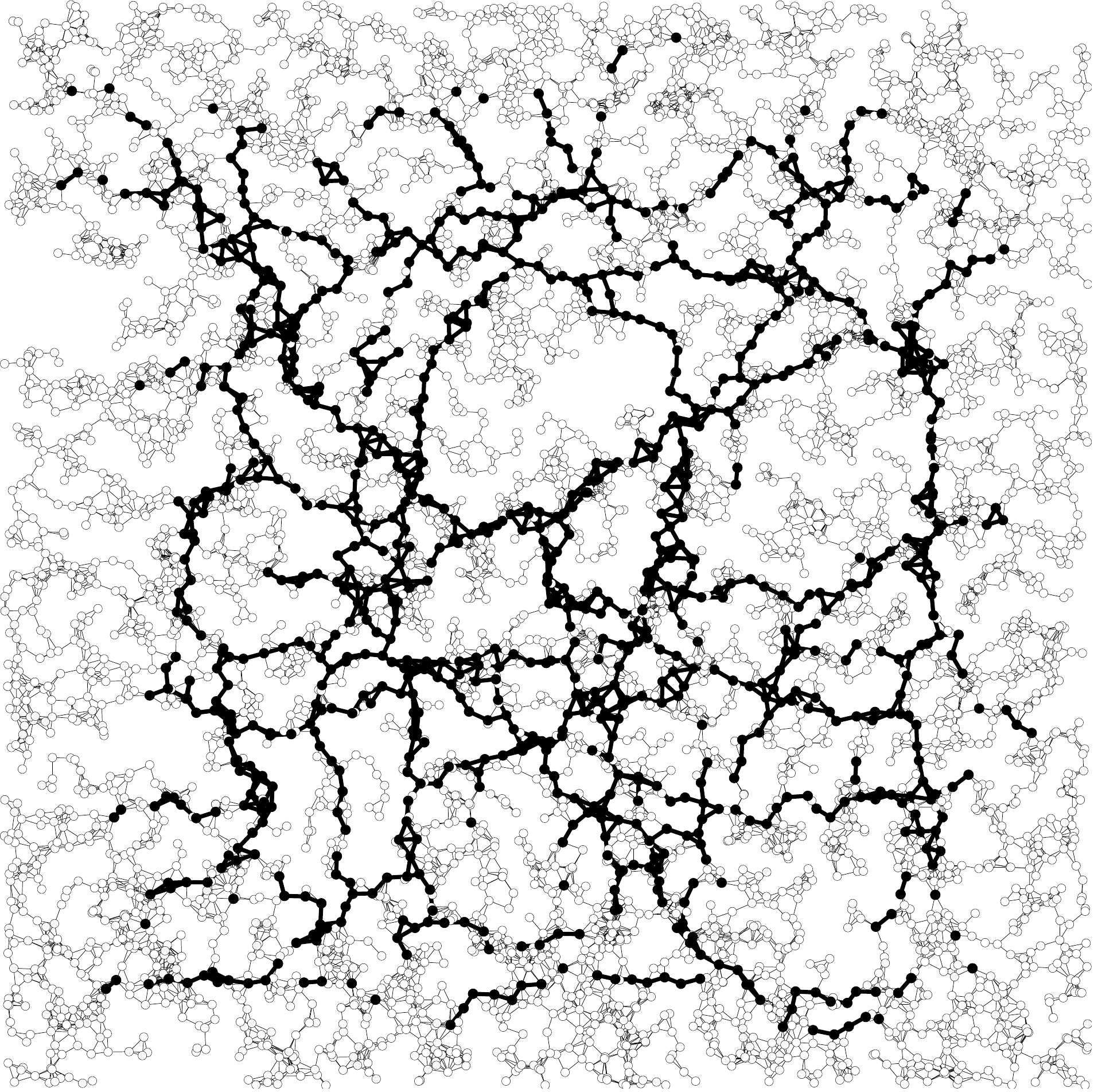}}
\subfigure[]{\includegraphics[scale=0.55]{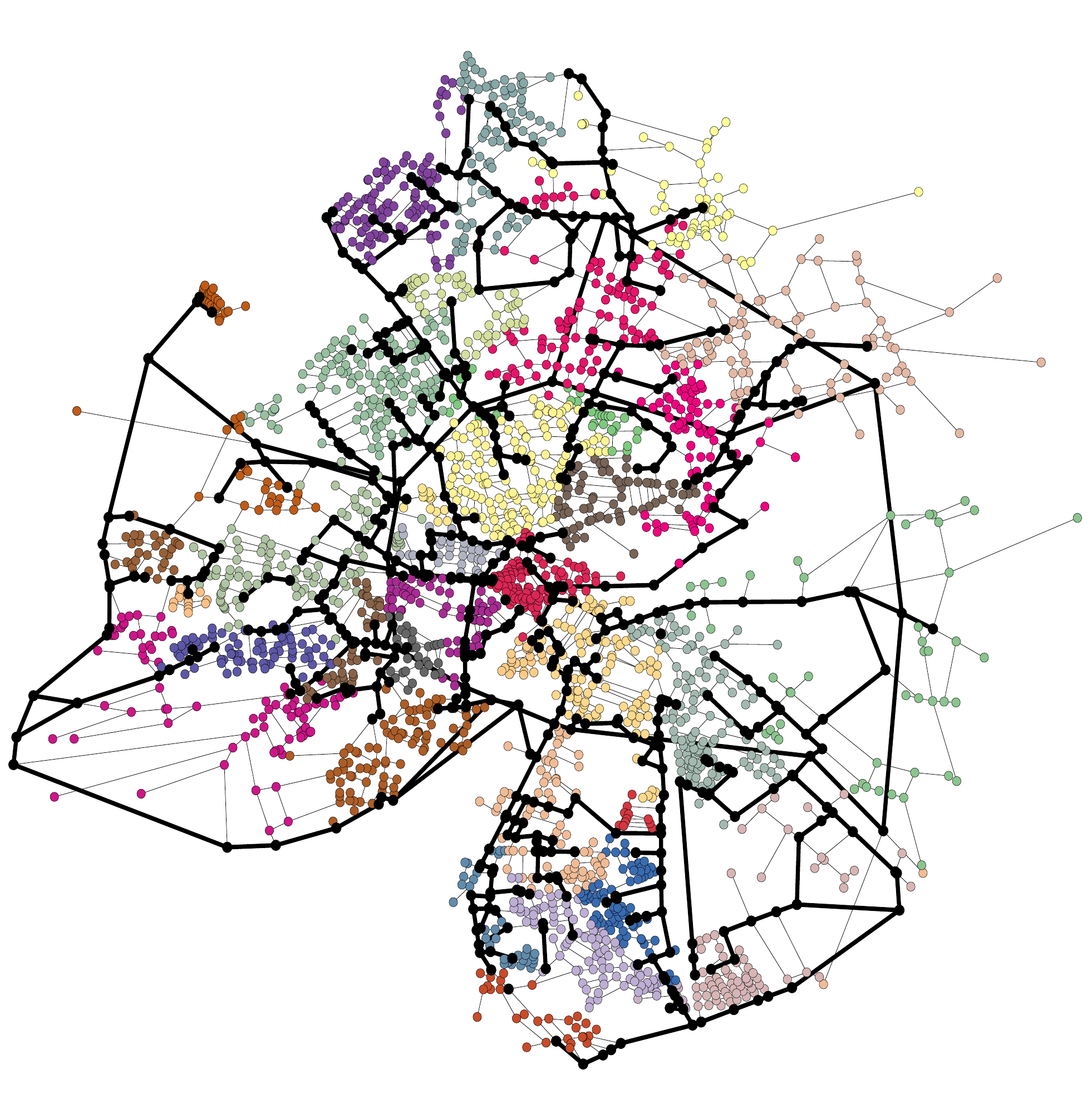}}
\caption{(Color online) Geographical networks, where black circles and highlighted edges show the preferential paths. Item (a) represents the
network of Waxman model, item (b) represents the network of the random geometric model, and item (c) represents the network of Oldenburg streets, where each
color represents a different community.}
\label{fig.networks}
\end{figure*}

In order to illustrate the betweenness paths, we simulated the Waxman network with 1707 vertices and average degree 2.20
(The network is shown in Figure \ref{fig.networks}(a)). The random geometric graph has 4883 vertices and average degree 6.18
(as can be shown in Figure \ref{fig.networks}(b)). The street network from San Joaquin has 14503 vertices and average degree 2.75,
and the network representing Oldenburg streets has 2873 vertices and average degree 2.57.
The Oldenburg network is shown Figure~\ref{fig.networks}(c), where
each color represents a different community. The San Joaquin network is not shown in Figure~\ref{fig.networks} because the network is very large, which hinders
a proper visualization of its paths and communities.

For each network we computed the betweenness centrality and highlighted the 20\% of vertices with largest betweenness values, which defines the characteristic
betweenness paths. The characteristic betweenness paths occurred in all examples, which are the characteristic vertices. By visual inspection, in each network
the characteristic vertices seem to define a backbone structure having many two-degree vertices. Such backbone is formed by what we call the \emph{characteristic
betweenness paths}. In street networks this result can indicate a tendency that the volume of cars is bigger in some streets than in others if we consider that
all drivers are using the shortest paths for the formation of high flow streets.

In every example we could observe the characteristic betweenness paths and it was possible to identify more than one component with high betweenness centrality
in the same network. The component sizes were different for each network. In the random geometric model, the giant component represents 85\% of all
vertices in the characteristic path. In the other networks the giant component represents more than 98\% of all vertices in the paths. 

We analyze the coverage of the characteristic paths. In order to do so, we measured the shortest distance between each secondary vertex $i$ and preferential
paths (represented as $d_i$), because the shorter the distances are, the better are the paths coverage. In order to provide a fair comparison between different
networks, the calculated distances are normalized by the  average path length, $l$, of each network. Therefore, we define the normalized shortest path distance,
$\Theta_i$, as
\[
\Theta_i = \frac{d_i}{l}.
\]

\begin{figure}
\center
\subfigure[]{\includegraphics[scale=0.52]{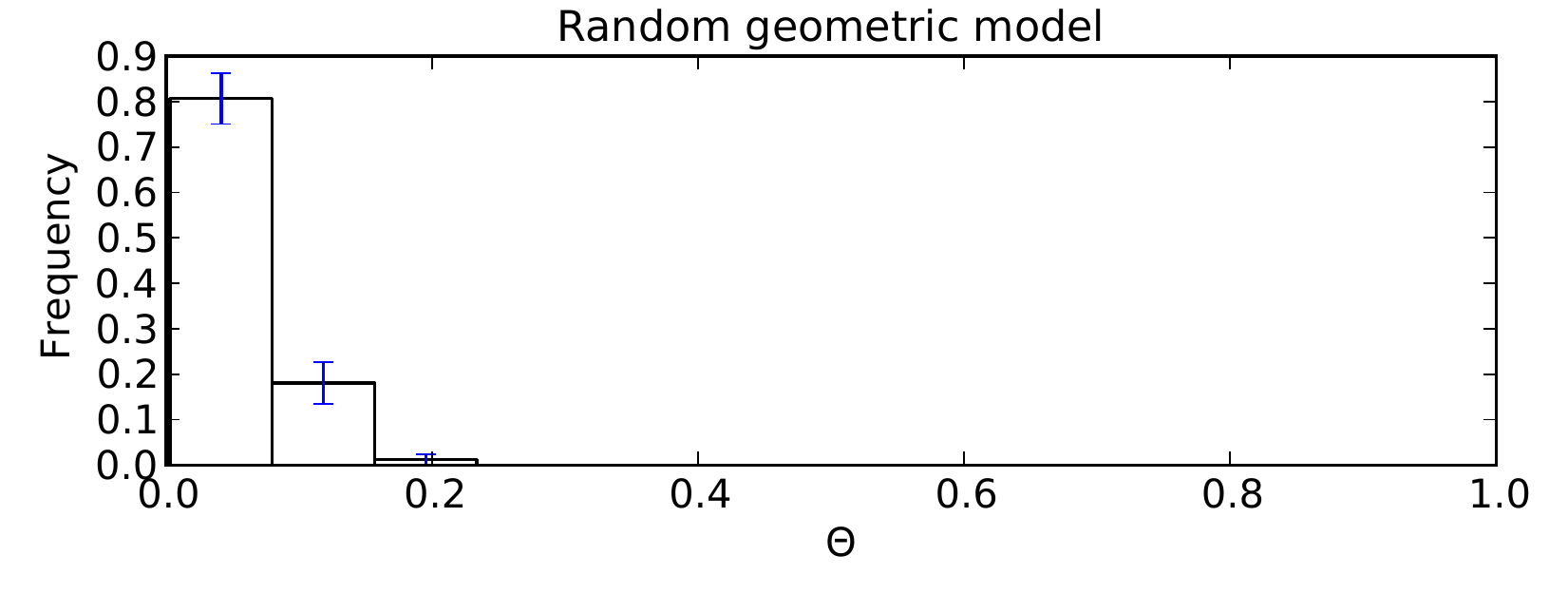}}
\subfigure[]{\includegraphics[scale=0.52]{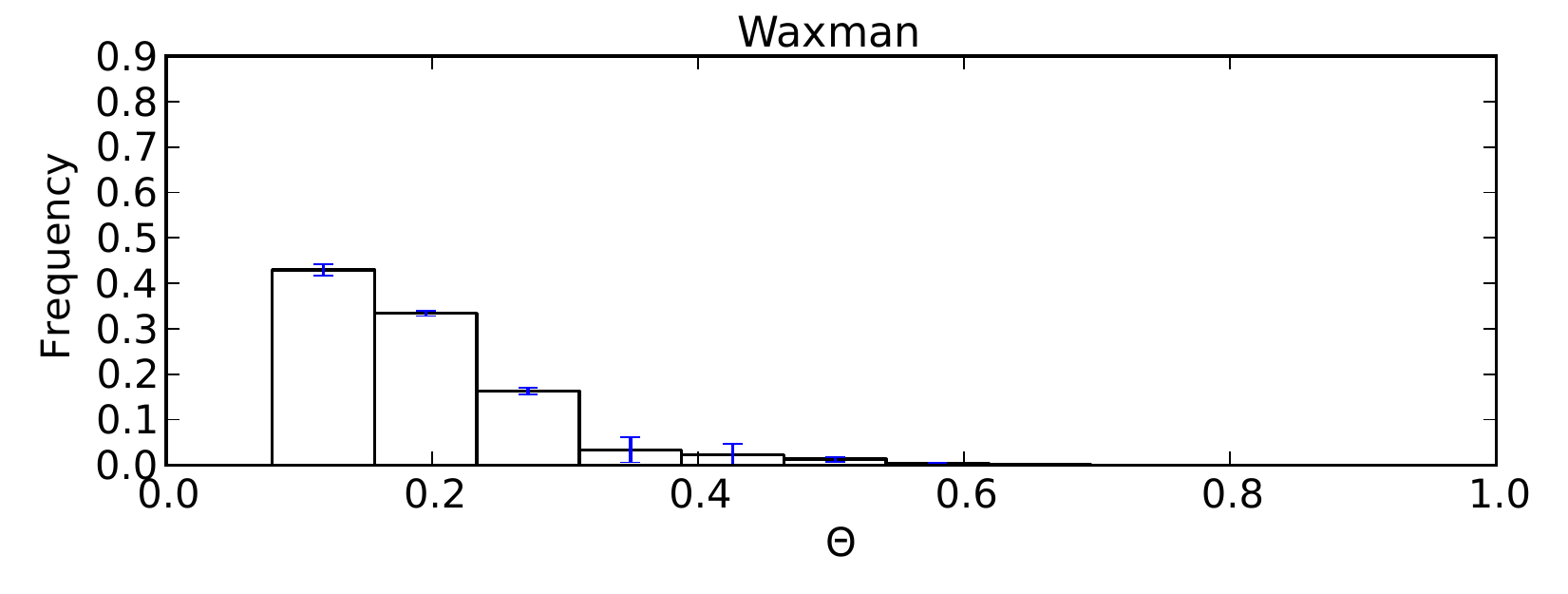}}
\subfigure[]{\includegraphics[scale=0.52]{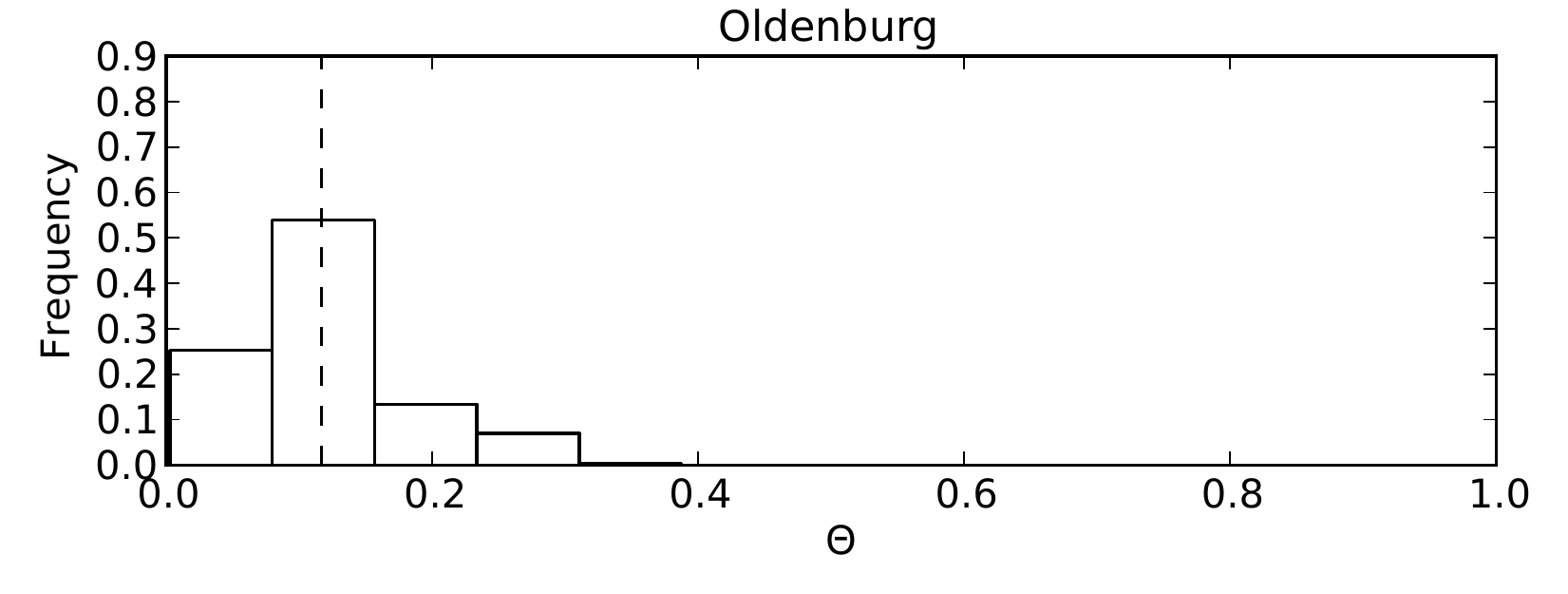}}
\subfigure[]{\includegraphics[scale=0.52]{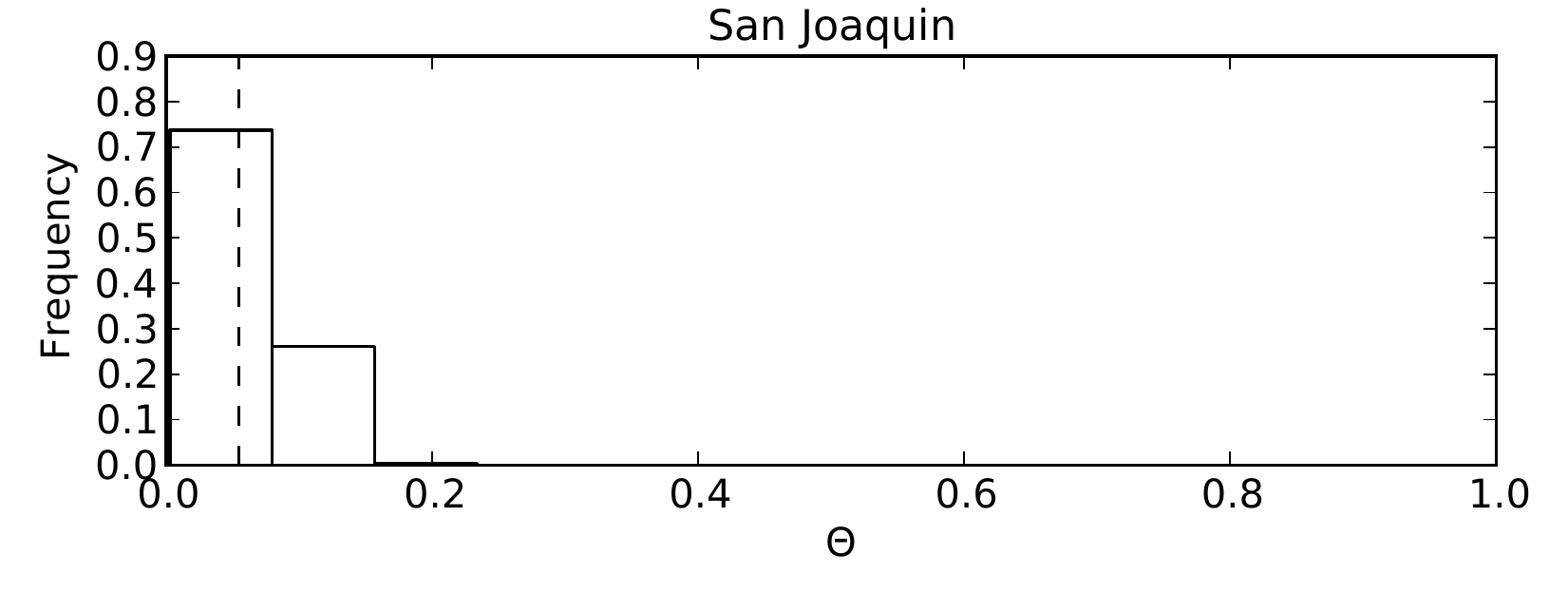}}
\caption{The histograms show the normalized shortest path distance, $\Theta$,  between each secondary vertex and preferential paths. We consider
 the (a) random geometric model, (b) Waxman model, (c) the Oldenburg network and (d) the San Joaquin network.  The dashed lines represent the coverage 
 and error bars in items (a) and (b) shows the standard deviation.}
\label{fig.histVertices}
\end{figure}

We generate $60$ samples of each model. In this case, we fit the model variables in order to compare with the Oldenburg network (modularity 0.92). The samples
of Waxman model have $2976.70\pm16.61$ vertices, average degree $2.71\pm0.04$, and modularity $0.79\pm0.01$. The samples of random geographic model
have $2892.33\pm58.46$ vertices, average degree $4.50\pm0.09$, and modularity $0.97\pm0.001$. In order to do a comparison
considering the statistical variations of the models, these results are shown as histograms in Figure~\ref{fig.histVertices}. We observed that the
fraction of vertices with the fewer distance (first bin in the histogram) is longer in the random geometric model than the Waxman model comparing the two
models, we observe that secondary vertices tend to be closer to characteristic paths in the random geometric graph model, but in this network the longest
distance is small than that case (as can be seen in Figures~\ref{fig.histVertices}(a)~and~(b)). Furthermore, the standard deviation is smaller in two networks
for both models, showing that this effect is not a statistical artifact caused by fluctuation. The shortest distances were also analyzed for the street networks
(as can be seen in Figures~\ref{fig.histVertices}(c)~and~(d)). The dashed lines represent the average values of $\Theta$,  which we call the coverage of the
 characteristic paths. The coverage of Oldenburg is larger than 
for San Joaquin, which may be a consequence of the higher complexity of Oldenburg's topology~\cite{Brinkhoff2002Cities}.


\section{Characteristic paths model}
\label{model}
In order to better understand the emergence of characteristic paths in geographic networks, we define a model able to generate networks with a simple
lattice-like topology, but still containing a well-defined backbone structure. The initial structure of the network is a 2D regular lattice,
where the vertices are organized in lines and equally spaced columns. First we defined the size, $L$, of the network (number of lines and columns) and two variables, the maximum 
distance to connect two vertices as $d_{MAX}$ and a probability $x$ to remove randomly the edges for each vertex. The algorithm follows three steps: 1- The vertices are organized as a 
square grid with toroidal boundary condition; 2- For each pair of vertices $i$ and $j$  when the distance $d_{ij}$ is less or equal than $d_{MAX}$ an edge connecting $i$
to $j$ is created; 3- Each edge is removed with probability $x$.

In order to visualize the betweenness paths in the model presented in this section, we created a network using the parameter $L = 200$, so that the network
has $N = 40000$. The connectivity range was set to $d_{MAX} = \sqrt{2}$. Therefore, each vertex of the network has degree 8. We randomly remove the
connections with probability $x = 0.65$.

\begin{figure*}
\center
\subfigure[]{\includegraphics[scale=0.8]{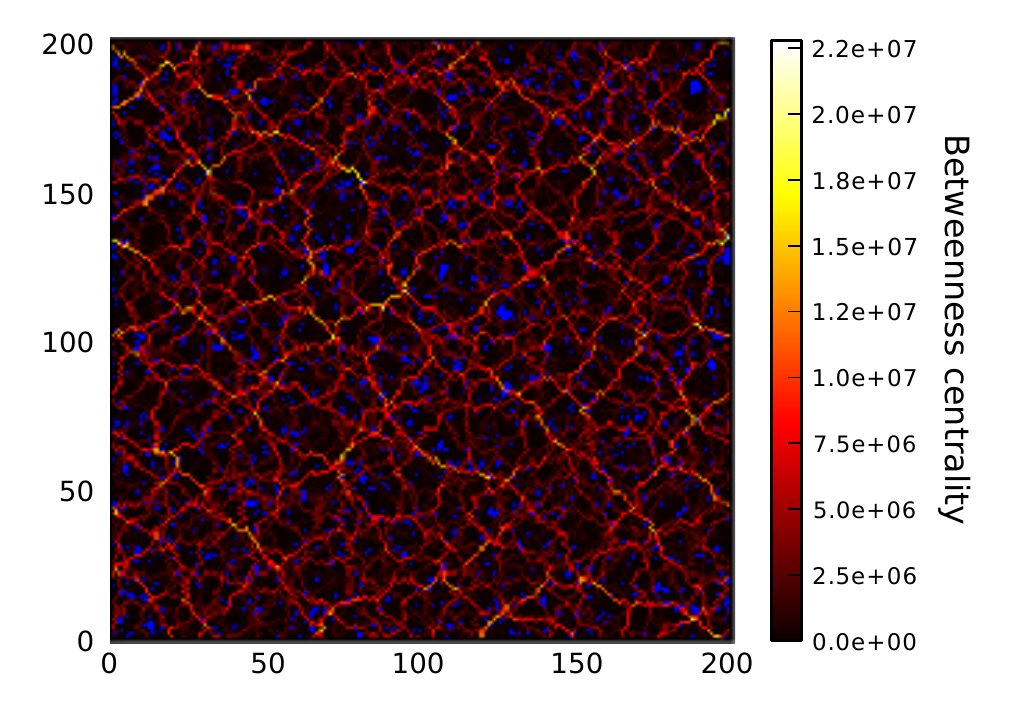}}
\subfigure[]{\includegraphics[scale=0.8]{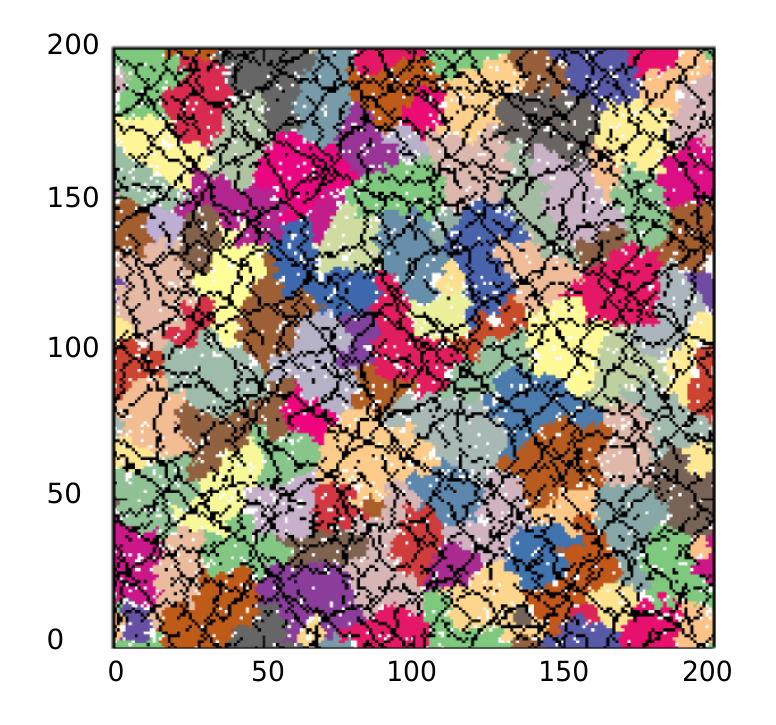}}
\caption{(Color online) Item (a) presents a heat map showing the betweenness centrality of vertices belonging to a simple  realization of our model.  The color blue shows
 regions without vertices.  In item (b) each color represents a different community, some distinct, non-adjacent, communities are associated to the same color due to color 
 palette limitations.. The color black shows the preferential paths. The color white shows regions  without vertices.}
\label{fig.colormaps}
\end{figure*}

We measured the betweenness centrality for each network vertex and the results can be seen in Figure~\ref{fig.colormaps}(a). We can observe
vertices with higher values (characteristic vertices) generating preferential paths in the network, in a similar manner as main streets. 
Our hypothesis is that the network communities are inducing the formation of betweenness paths through connections between them. 
In Figure~\ref{fig.colormaps}(b) we show the network communities and characteristic paths (indicated in black). The result suggests that the paths are going through communities. 

To compare the characteristic betweenness paths displayed by our network model and real networks, we executed the same tests as in the previous section.
In order to infer the statistical variation of the model, we created $60$ network samples. We used the parameter $L = 56$ and we randomly removed
the connections with probability $x = 0.7$. These parameters were set so that the generated networks have a size and average degree as close as possible to compare with the Oldenburg network. The network measurements are: number of vertices $2829.95\pm21.34$, average degree $2.59\pm0.03$, and modularity $0.94\pm0.002$. 

\begin{figure}
\center
\includegraphics[scale=0.52]{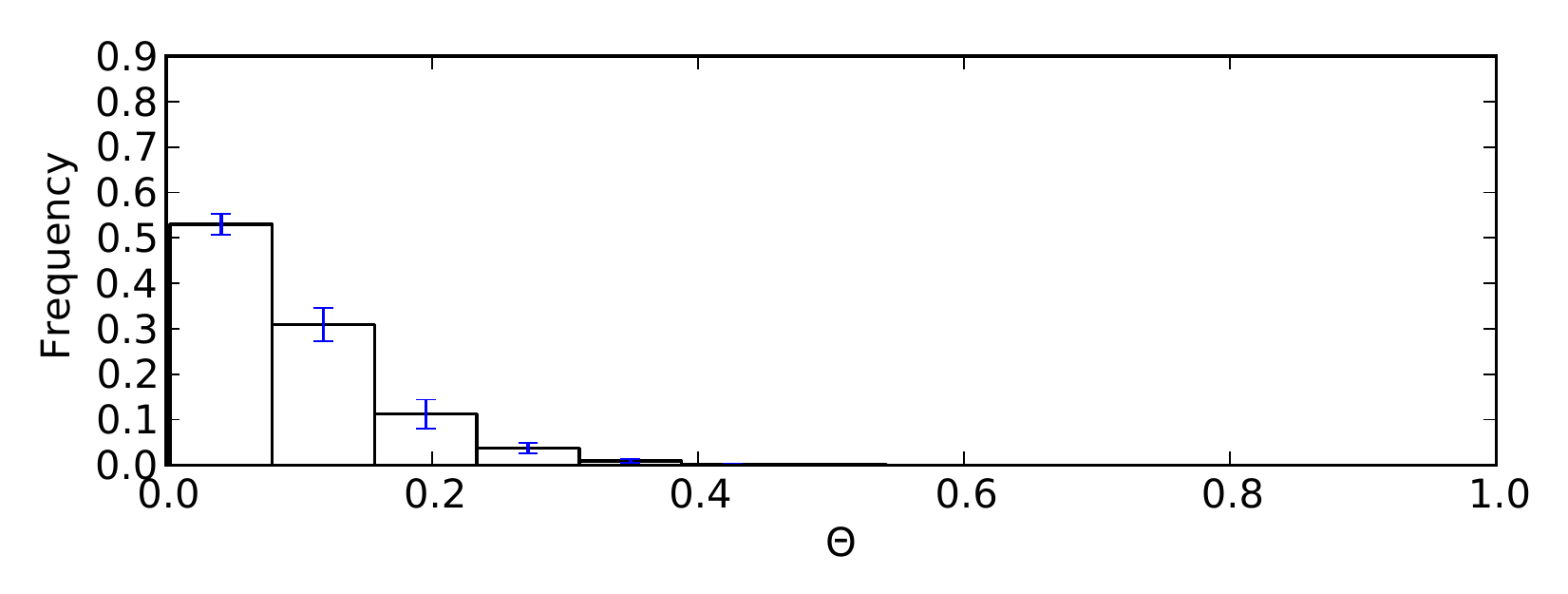}
\caption{Histogram of normalized shortest path distance, $\Theta$, between each secondary vertex and preferential betweenness paths.}
\label{fig.histGeographic}
\end{figure}

Our model has similarities and differences from the street networks and the other models. The giant componet represents
$(85\pm8)\%$ of all characteristic vertices. Regarding the shortest distance  between secondary vertices and preferential paths (Figure~\ref{fig.histGeographic}), 
the secondary vertices tend to be more distant from the backbone than in the Oldenburg and Waxman networks. 

\begin{figure}
\center
\includegraphics[scale=0.52]{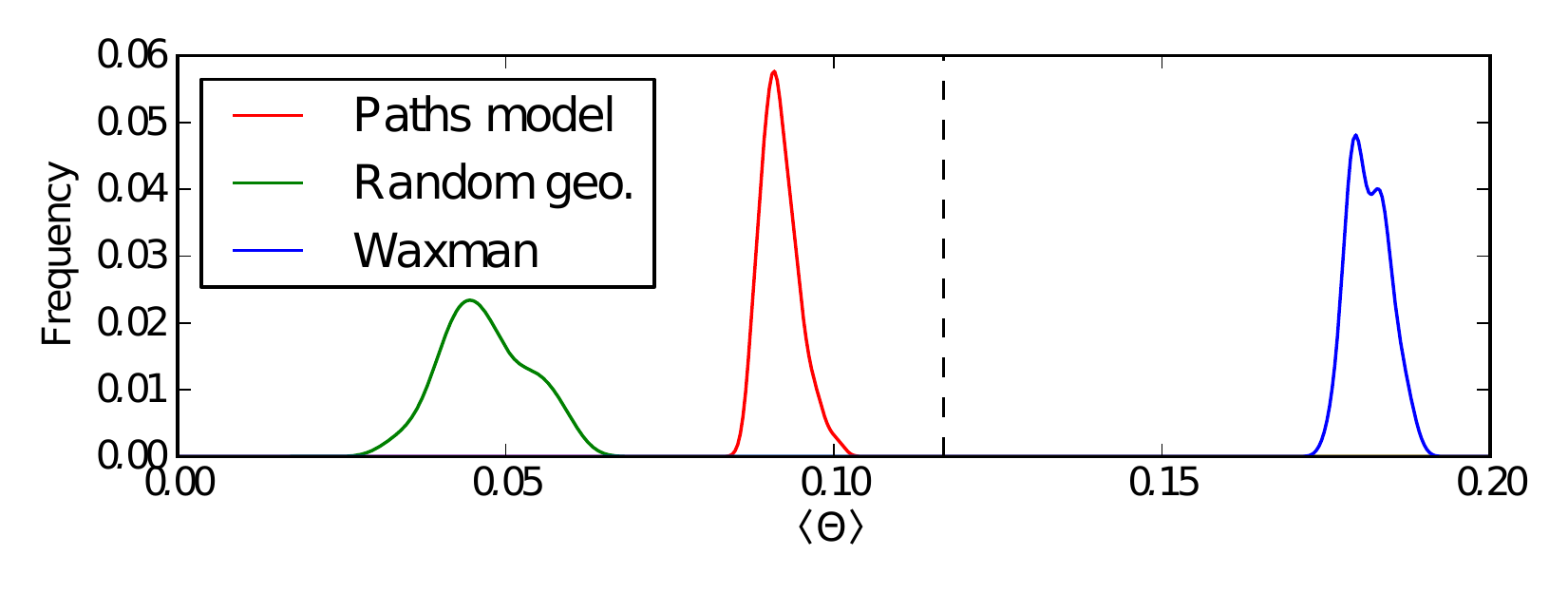}
\caption{Histogram of the coverage for $60$ generated networks from each considered model. The dashed line represents the
coverage in Oldenburg's network.}
\label{fig.histModels}
\end{figure}

We compared the average of shortest distance between the secondary vertices and the preferential betweenness paths in network models and the Oldenburg 
network. For each generated network we calculate coverage, which we represent by  $\langle \Theta \rangle$, and in Figure~\ref{fig.histModels} we show the
distribution of  $\langle \Theta \rangle$ for the generated networks. The dashed line shows the coverage for Oldenburg. The coverage for 
Oldenburg is closest to the measured values for the characteristic path model. Besides, the standard deviation  is low, which confirms that the betweenness paths have a similar structure in all generated networks. Therefore, in the following we further analyze the betweenness paths using  a single realization of our model.  

\begin{figure*}
\center
\subfigure[]{\includegraphics[scale=0.57]{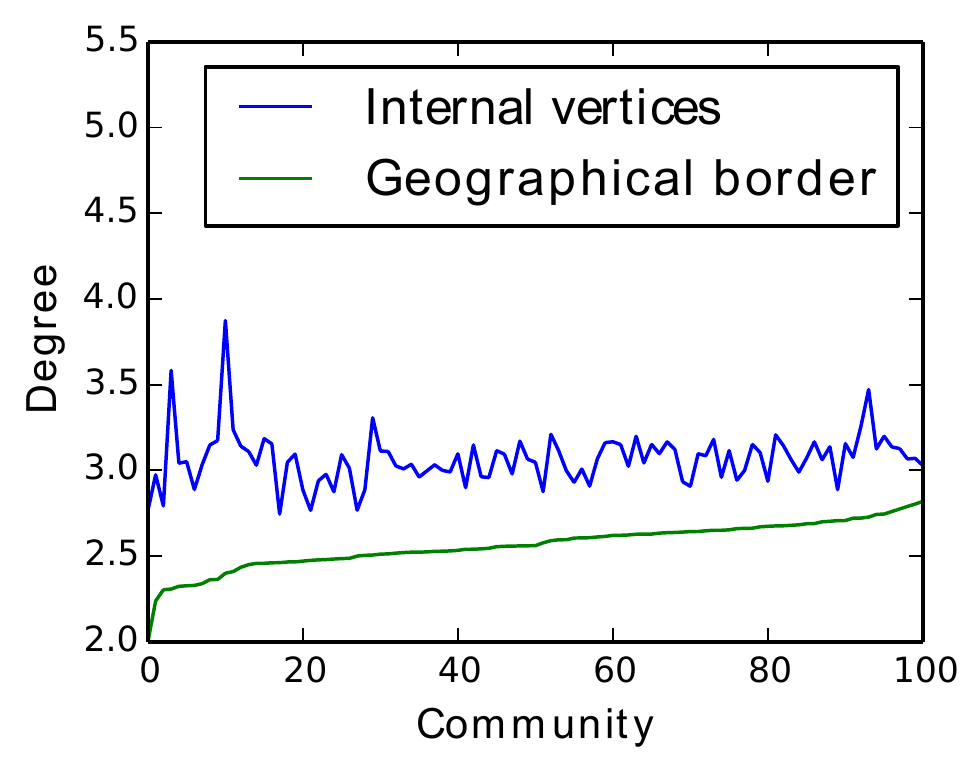}}
\subfigure[]{\includegraphics[scale=0.57]{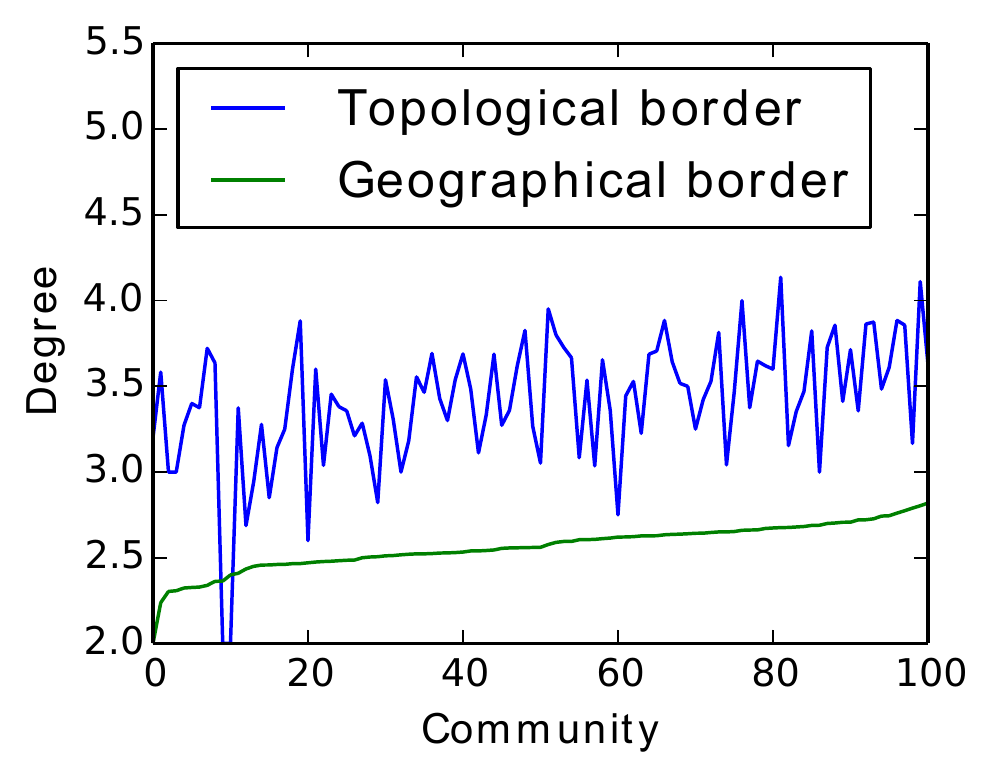}}
\subfigure[]{\includegraphics[scale=0.57]{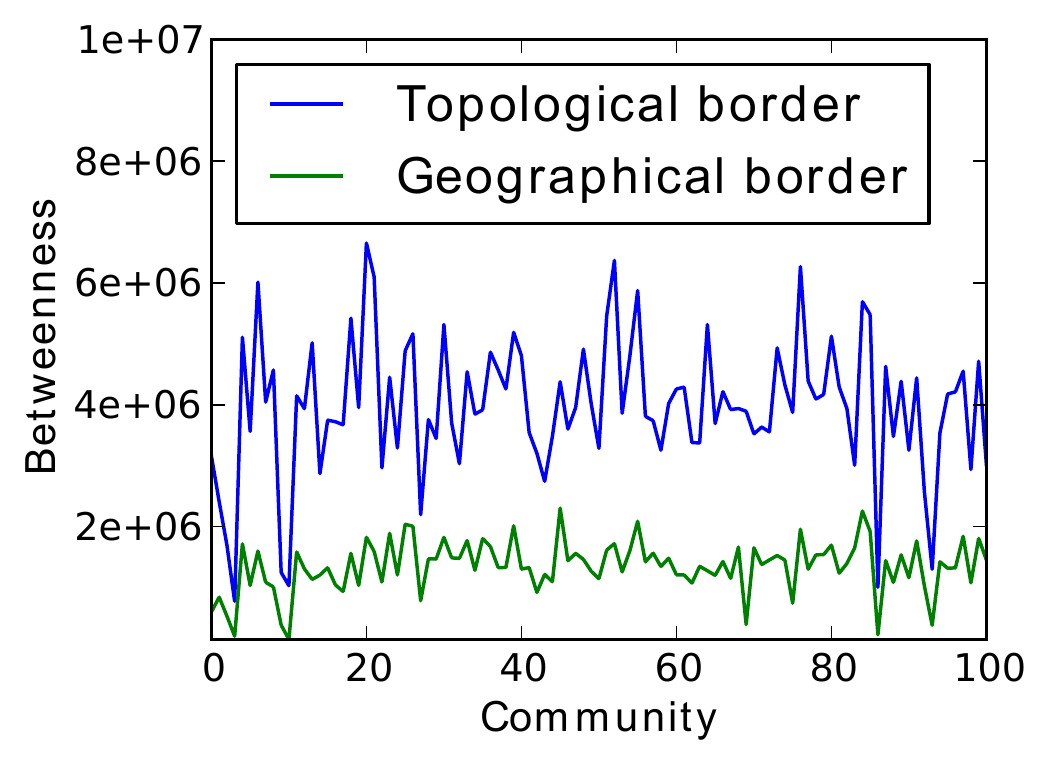}}
\caption{(Color online) Image (a) shows the degree comparison between internal vertices and border community vertices, image (b) shows the degree
comparison between topological and geographical border, and image (c) shows the betweenness centrality comparison between topological and geographical
border. The community indexes were ordered according to the degree of internal community vertices.}
\label{fig.2}
\end{figure*}

To quantify our network communities in terms of their respective degree and betweenness centrality, we considered two types of vertices, the geographical border
community vertices and the internal community vertices. The first type represents vertices that have at least one neighbor in the grid belonging to a different
community, no matter if they are connected or not. The second type represents a vertex that is not a geographical border. In order to define each vertex type, we 
considered all eight geographical neighbors, of the vertices. In some analyses we also considered the topological border community vertices, which are the vertices 
connected with one or more  vertices from   different community. We generated a network with $L=200$ and average degree $2.92$ according to our model. 
Using this network  we computed three different  comparisons. In the first, we compare the average degree between internal vertices and border community vertices
 (as can be seen in Figure~\ref{fig.2}(a)).  In this comparison it is possible to observe that the average degree in the internal community vertices is higher than the border community vertices. Thereby, we can affirm that the vertex degree is influencing how the community is formed.

In order to describe how the vertices are connecting communities, we measured the average degree in the geographical community border and in the topological
community border, represented in Figure \ref{fig.2}(b). In 98\% of the communities, the average degree of vertices belonging to the geographical border is lower
than for vertices contained in the topological border. 



We made a similar comparison, between geographic and topological borders, but now measuring the betweenness centrality instead of the degree (as can be seen in
Figure~\ref{fig.2}(c)). This analysis can show if connections between communities (topological border) are generating a characteristic betweenness paths.
Figure \ref{fig.2} shows that the average geographical border betweenness is lower than the topological border betweenness for all communities. Thus, in
average, the vertices where the paths are crossing the community borders have higher betweenness centrality and degree than the geographical borders.
Some vertices of characteristic paths are part of the topological border. In other words, they are connecting the communities.

\begin{figure}
\center
\includegraphics[scale=0.35]{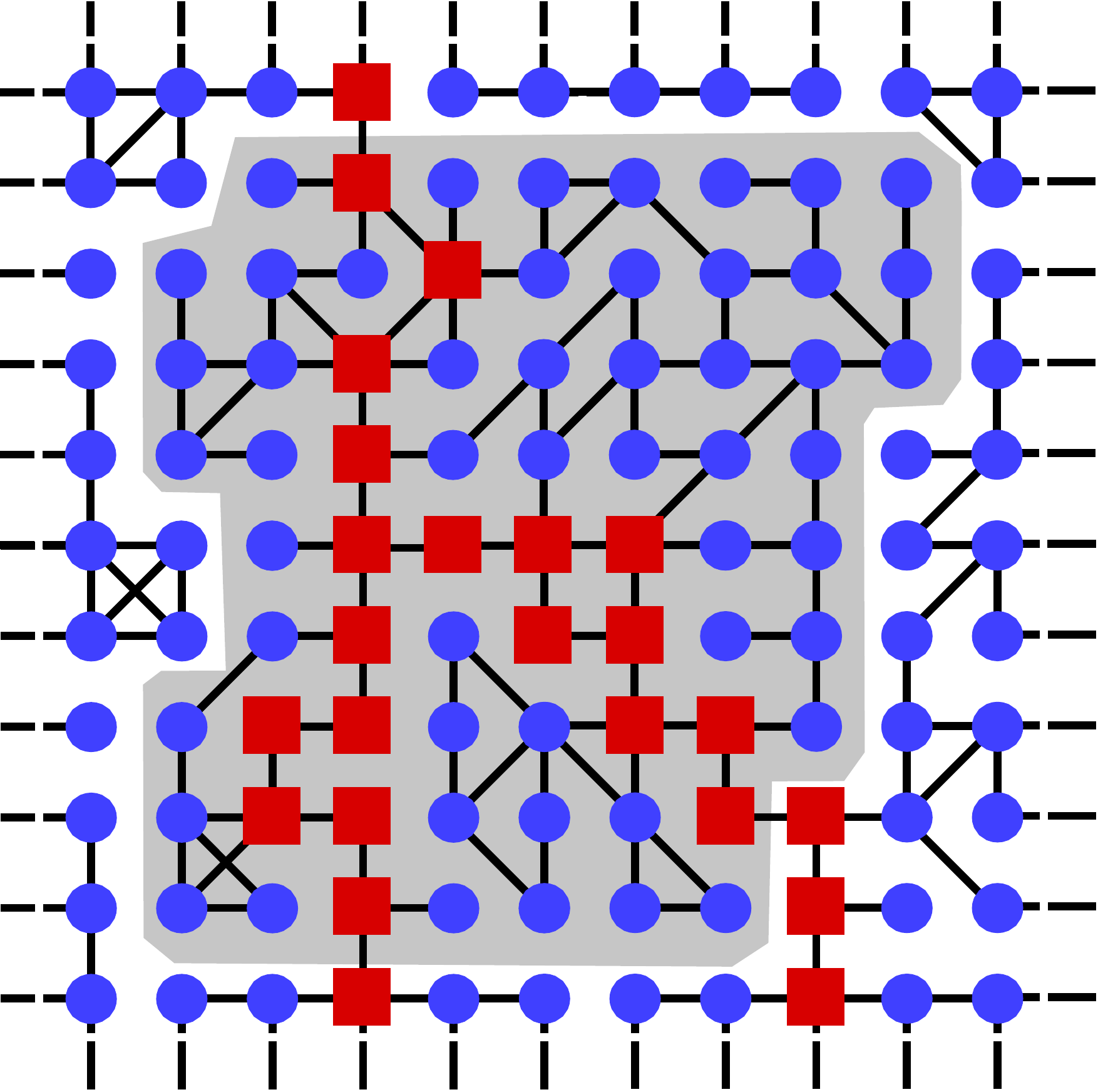}
\caption{(Color online) The highlighted region represents a community and the red squares represent characteristic vertices crossing the
community.}
\label{fig.3}
\end{figure}

The results indicate that the paths represent routes of efficient communication between the communities. This idea is illustrated in
Figure~\ref{fig.3}. Vertices included in the highlighted gray region belong to the same community. The community has three vertices on its topological border and the
shortest paths connecting them, shown in red, are creating the preferential paths with higher betweenness centrality.  

\begin{figure}
\center
\includegraphics[scale=0.45]{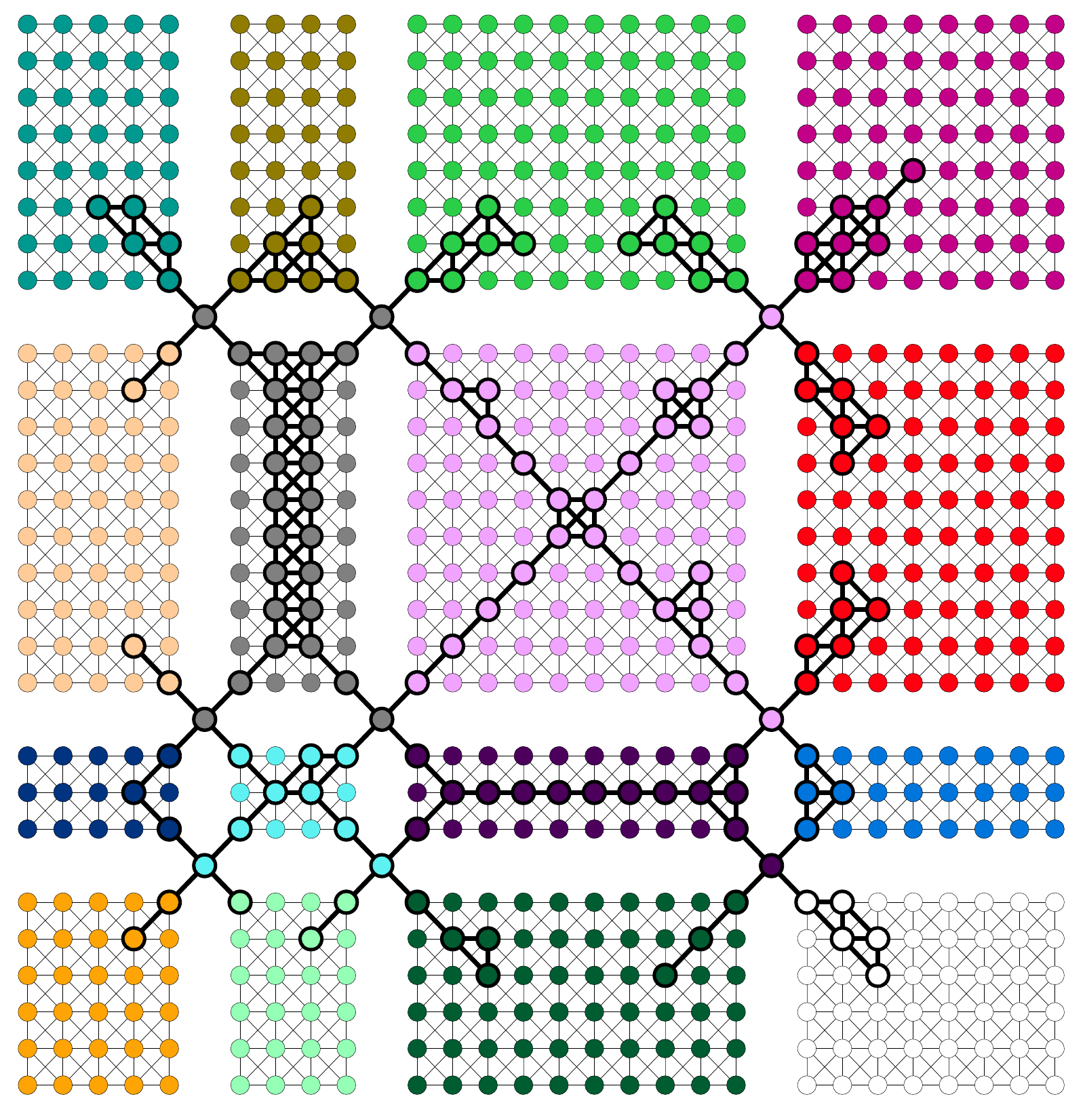}
\caption{(Color online) Network created by connecting lattice-like communities. Each community is represented by a different color
and the characteristic paths are highlighted.}
\label{fig.3.1}
\end{figure}

In order to validate these results, we created an artificial geographic network with fixed communities. In this network, each community is a regular lattice, 
where the vertices are connected with their eight nearest neighbors, and the communities are interconnected through some selected vertices. Figure~\ref{fig.3.1} 
shows the network and highlight the highest values of betweenness centrality (the characteristic vertices). In this test, it is possible to observe
that the characteristic paths are connecting the communities. Besides, through this model, it is  possible to note that the paths are not created at random, that is, 
their emergence have a well-defined mechanism.

\begin{figure}
\center
\includegraphics[scale=0.7]{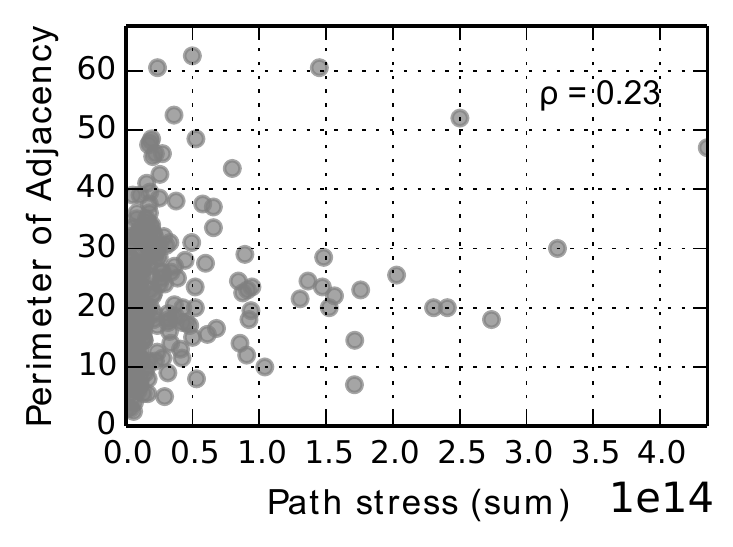}
\caption{Scatter plot showing for each community the summation of all stress connecting vertices from a community to another against the perimeter of
adjacency. The Pearson correlation ($\rho$) is shown inside the plot.}
\label{fig.4}
\end{figure}

It is also interesting to analyze if the number of shortest paths connecting pairs of communities are related to the sizes of the communities geographical border.
In order to do so, we plot the perimeter of adjacency between communities against the sum of the vertex stress connecting such communities, which can be seen in
Figure~\ref{fig.4}. Besides, we calculated the Pearson correlation coefficient between the measurements, obtaining a value of 0.23. Thus, these measurements are uncorrelated.

\section{Community adjacency network}
\label{community}

\begin{figure}
\center
\includegraphics[scale=0.56]{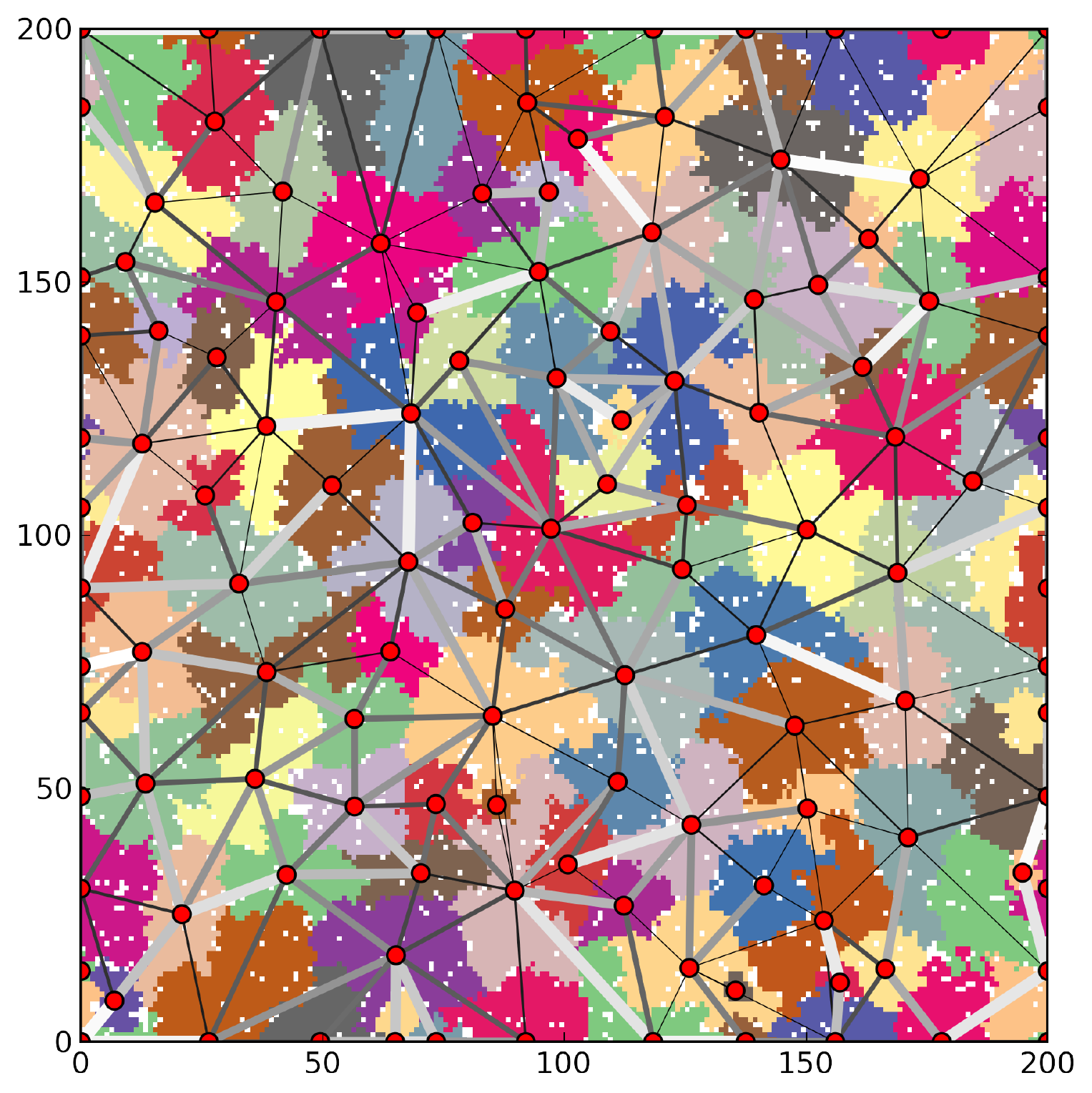}
\caption{(Color online) Heat map representing the community adjacency network from communities, where the color of edges represent the weight (black is zero and
white is one).}
\label{fig.5}
\end{figure}

In order to observe the relationship between communities and their importance, considering geographical characteristics, we created a network based on the
topological adjacency between communities of our characteristic paths model. Through this model it is possible to compare the geographical characteristics
(e.g., the community area) with topological characteristics, such as the degree of a vertex, which represents the number of communities connected with a
given community. In this network, each vertex represents a community. A weight equal to $SC_i + SC_j$ (defined in Equation~2)  is associated to each edge
connecting communities $i$ and $j$. The resulting network is undirected. In Figure~\ref{fig.5} we show a typical realization of this network. 
It is possible to observe small communities having strong connections, indicating that betweenness paths are also present in such small communities.
On the other hand, there are big communities with weak connections, showing that few preferential paths are crossing them.

\begin{figure*}
\center
\subfigure[]{\includegraphics[scale=0.7]{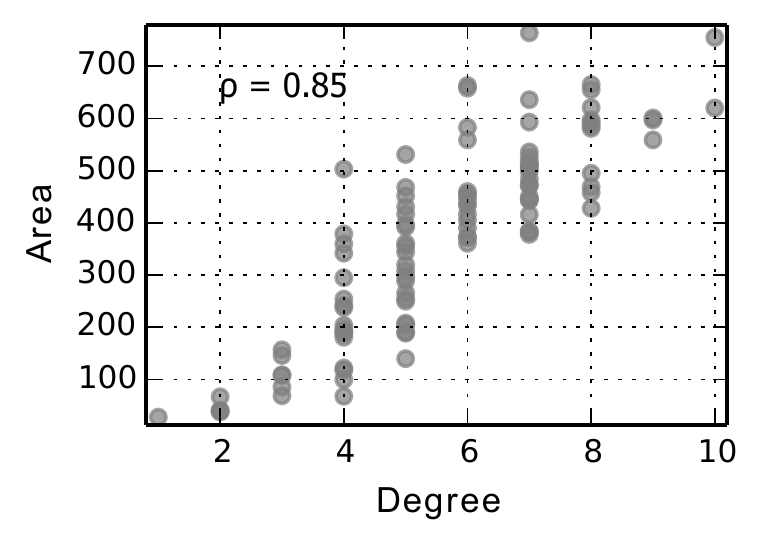}}
\subfigure[]{\includegraphics[scale=0.7]{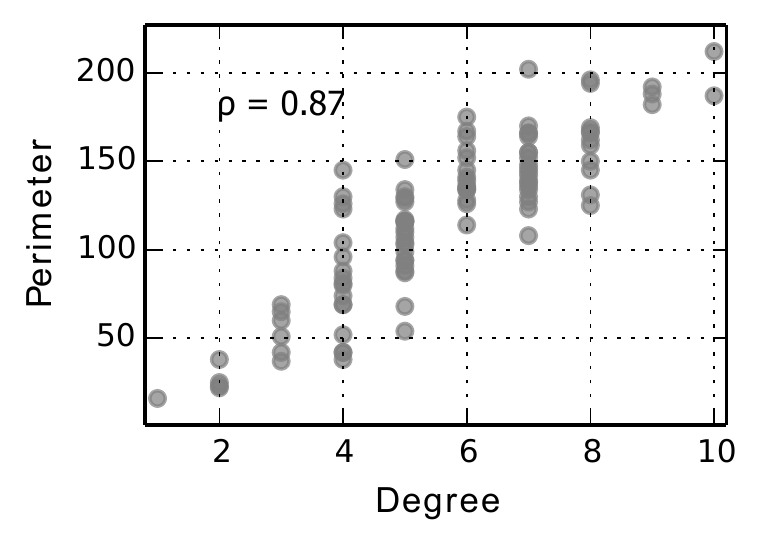}}
\subfigure[]{\includegraphics[scale=0.7]{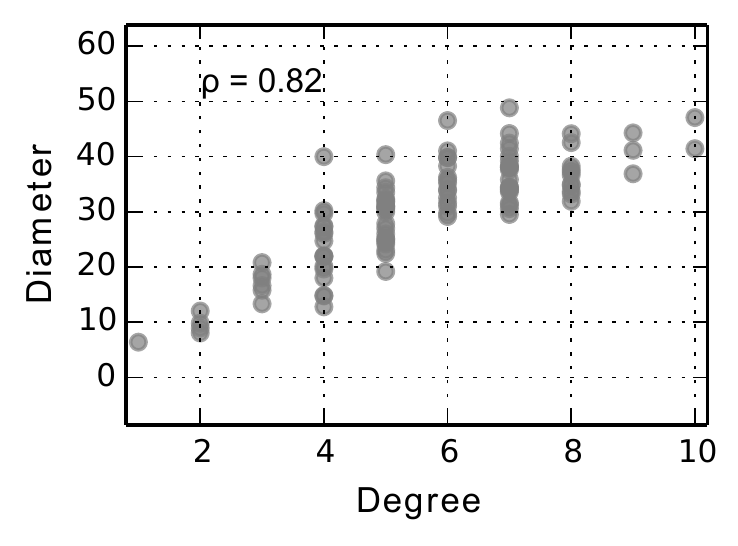}}
\subfigure[]{\includegraphics[scale=0.7]{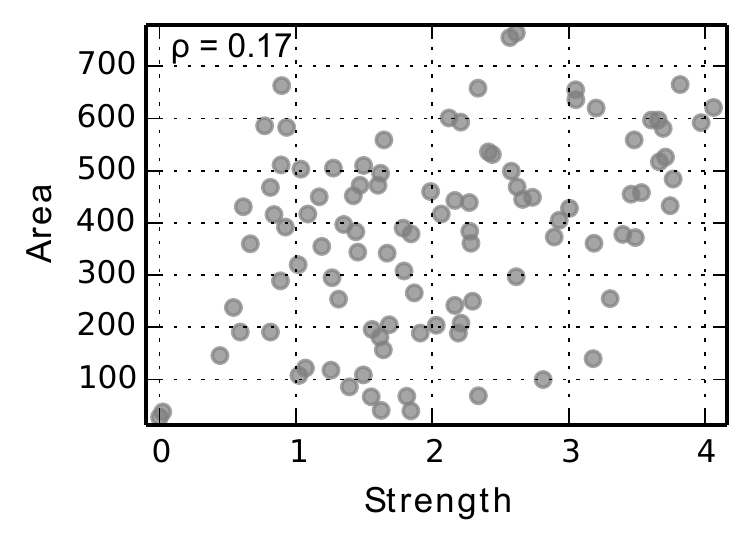}}
\subfigure[]{\includegraphics[scale=0.7]{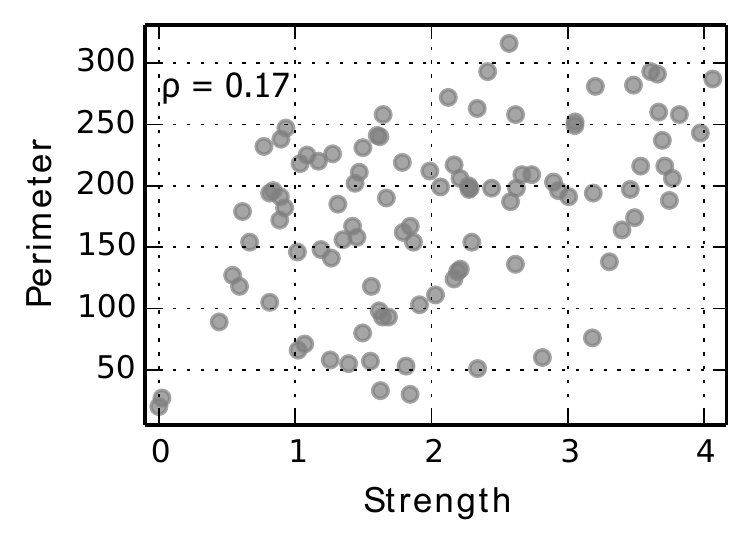}}
\subfigure[]{\includegraphics[scale=0.7]{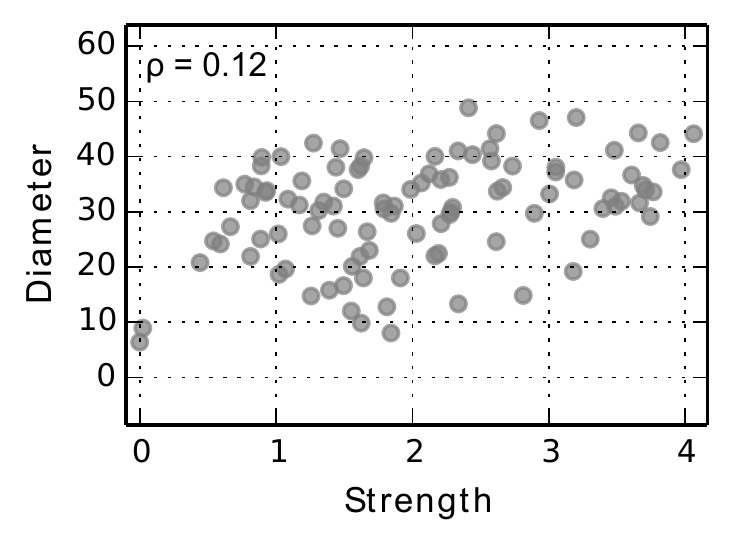}}
\caption{Scatter plots comparing geographical and topological measurements in the community networks. The items (a), (b), and (c) represent the unweighted
degree against geographical characteristics and the items (d), (e), and (f) represent the strength against geographical characteristics. The Pearson correlation,
$\rho$, is shown inside each plot.}
\label{fig.6}
\end{figure*}

We compared local topology characteristics of the community adjacency network with geometrical characteristics (the scatter plots can be seen in
Figure~\ref{fig.6}). Figures~\ref{fig.6}(a),~(b)~and~(c) show the degree as a function of area, perimeter and diameter, respectively and Figures~\ref{fig.6}(d),
(e)~and~(f) show the same measures using the vertex strength instead of the degree. The degree is highly correlated with the
geographical characteristics of communities. On the other hand, the strength has little relationship with the communities geometry. This happens because the
 strength is uncorrelated with the geographical characteristics of communities.

\section{Conclusion}
\label{conclusion}

The structure of a geographical network has a direct impact on how the system can be navigated. Our analysis revealed that betweenness paths are a common
characteristic of geographical networks,  since they can be found in geographical models as well as real world networks. We characterized the betweenness paths
in terms of the distance between each network secondary vertex and the closest betweenness path. Such measurement can be related to the covering provided by the paths to reach the entire network. We showed that in both the geographical models and the street networks the betweenness paths provide a good covering of the system.  
In order to better understand the establishment of the characteristic paths, we proposed a model to generate geographical networks having a simple topology but
that can still originate such paths.

It was found that the characteristic betweenness paths have a strong relationship with the presence of well-defined communities in the networks. We compared the
vertices degrees and betweenness centrality among geographical and topological borders of the communities, as well as the internal vertices of the communities.
This analysis revealed that the betweenness characteristic paths are usually  induced by the communities, in the sense that such paths provide a communication 
between entry and exit points of the communities.

The number of shortest paths passing through the topological border of adjacent communities seems to be independent of the  geometry of the interface between
the communities. This was  revealed by the absence of correlation between the total stress of vertices connecting adjacent communities and the perimeter of the
adjacency. In order to better understand the relationship between communities and characteristic betweenness paths, we defined a new structure based on the
vertices betweenness and the community organization of the network being studied, which we called community network. By characterizing this meta structure, we
concluded that geographical characteristics can influence the connectivity between communities, which takes place along betweenness characteristic paths.

It remains an open question if other, non-geographical, network models can also present such characteristic betweenness paths. Also, the proper impact that
such paths may have on a dynamics taking place on the system is also an interesting aspect for future studies.

\section{acknowledgments}

Henrique Ferraz de Arruda thanks CNPq (Grant no. 132363/2013-5) and CAPES for financial support. Cesar Henrique Comin acknowledges
FAPESP (Grant no. 11/22639-8) for financial support. Luciano da Fontoura Costa is grateful to CNPq (Grant no. 307333/2013-2), FAPESP (Grants no. 11/50761-2), and NAP-PRP-USP for sponsorship. The authors thanks Filipi Nascimento Silva for fruitful discussions.  

\bibliography{manuscript}

\end{document}